\documentclass[a4paper]{spie}

\usepackage[latin1]{inputenc}
\usepackage{psfrag}
\usepackage{amssymb}
\usepackage{amsmath} 
\usepackage{epsf}
\usepackage{epic}
\usepackage{color}
\usepackage{cite}
\usepackage{amsfonts}
\usepackage{graphics}
\usepackage{multirow}

\title{Accurate measurement of Cn2 profile with Shack-Hartmann data}

\author{Juliette Voyez\supit{1}, Clélia Robert\supit{1}, Vincent
  Michau\supit{1}, Jean-Marc Conan\supit{1}, Thierry Fusco\supit{1}\\
  \supit{1}Onera, The French Aerospace Lab, Dép\textsuperscript{t} d'Optique
  Théorique et Appliquée, 92322 Châtillon, France \\
}
\authorinfo{Further author information: juliette.voyez@onera.fr}

\begin{document}

\maketitle

\begin{abstract}
The precise reconstruction of the turbulent volume is a key point in the
development of new-generation Adaptive Optics systems. We propose a new
$C_{n}^{2}$ profilometry method named CO-SLIDAR (COupled Slope and
scIntillation Detection And Ranging), that uses correlations of slopes and
scintillation indexes recorded on a Shack-Hartmann from two separated stars.
CO-SLIDAR leads to an accurate $C_{n}^{2}$ retrieval for both low and high altitude
layers. Here, we present an end-to-end simulation of the $C_{n}^{2}$ profile
measurement. Two Shack-Hartmann geometries are considered. The detection
noises are taken into account and a method to subtract the bias is proposed.
Results are compared to $C_{n}^{2}$ profiles obtained from correlations of slopes only
or correlations of scintillation indexes only. 
\end{abstract}
\vspace{10pt}

\section{Introduction}\label{sect:intro}

An accurate knowledge of the vertical distribution of the strength of
atmospheric optical turbulence is a crucial point in the development of
Adaptive Optics (AO) facilities. New Wide Field Adaptive Optics (WFAO) concepts
are investigated for the Extremely Large Telescopes (ELTs) in order to
increase the field of correction. The impact of the $C_{n}^{2}$ profile on the
performances of WFAO systems has been pointed out\cite{2010SPIE.7736E..17F}, leading
to the need for precise tomographic reconstruction of the turbulence volume.

CO-SLIDAR\cite{Vedrenne:07} is a new method to measure a high-resolution
$C_{n}^{2}$ profile. The principle is to use correlations of wavefront slopes and scintillation
indexes recorded with a Shack-Hartmann (SH) from a binary star. CO-SLIDAR combines sensitivity to ground and low altitude
layers, with correlations of slopes, like a SLODAR\cite{2002MNRAS.337..103W},
but it also provides sensitivity to weak high altitude layers, taking
advantage of correlations of scintillation indexes, like
SCIDAR\cite{1998PASP..110...86F} or MASS\cite{2003SPIE.4839..837K}. 

CO-SLIDAR first tests in simulation\cite{Vedrenne:07} and on-real
data\cite{2011arXiv1101.3924R} are encouraging. The extension of the method to
a single source has also been tested on infrared data\cite{Vedrenne:10}, in an
endo-atmospheric context. We are now looking towards a complete on-sky
validation of the concept. To prepare it, we perform an end-to-end simulation
of a $C_{n}^{2}$ profile measurement with CO-SLIDAR, in a concrete
astronomical case, with binary stars observed on a $1.5$-meter telescope,
taking into account realistic fluxes and detection noises.

In Section \ref{sect:problem}, we recall CO-SLIDAR theoretical background. In Section
\ref{sect:simu}, we set out the simulation of astronomical turbulent images
with two SH geometries. Section \ref{sect:data_processing} is dedicated to the
data processing. In Section \ref{sect:results}, $C_{n}^{2}$ reconstruction results are shown and commented. Our conclusions and
perspectives are exposed in Section$~$\ref{sect:conclusion}.

\section{Problem statement}\label{sect:problem}

The analytical formulation and notations are recalled here, for the
understanding of the problem statement and its inversion, to get an accurate measurement of
the $C_{n}^{2}$ profile with Shack-Hartmann data.
We consider two stars with separation $\mathbf{\theta}$ in the field of view (FOV).
The SH delivers a set of wavefront slopes and intensities per frame and per
star. For one star, we denote $\mathbf{s_{m}}\left(\mathbf{\alpha}\right)$ the slope measured in the
subaperture $m$, $\mathbf{\alpha}$ being the position of the star. It is a
bidimensionnal vector with two components $s^k_m(\mathbf{\alpha})$,
along the $k$ axis ($k \in \{x,y\}$), corresponding to the two ``tip'' and
``tilt'' directions. The star intensity is denoted $i_m(\mathbf{\alpha})$, and leads to
the scintillation index $\delta
i_m(\mathbf{\alpha})=\frac{i_m(\mathbf{\alpha})-o_m(\mathbf{\alpha})}{o_m(\mathbf{\alpha})}$
where $o_m(\mathbf{\alpha})$ is the time-averaged star intensity. 
Slope correlations $\langle s^k_m s^l_n \rangle(\mathbf{\theta}) $ and scintillation index correlations
$\langle \delta i_m \delta i_n \rangle(\mathbf{\theta})$ can be written as
$C_n^2(h)$ integrals weighted by functions denoted $W_{ss}^{kl}$ and $W_{ii}$
in the following expressions:
\begin{equation}
  \label{eq:corr_slo}
  \langle s^k_m s^l_n \rangle(\mathbf{\theta}) = \int_{0}^{+\infty}C_{n}^{2}\left(h\right)W_{ss}^{kl}\left(h,\mathbf{d}_{mn},\mathbf{\theta}\right)dh,
\end{equation}
\begin{equation}
  \label{eq:corr_sci}
  \langle \delta i_m \delta i_n \rangle(\mathbf{\theta}) = \int_{0}^{+\infty}C_{n}^{2}\left(h\right)W_{ii}\left(h,\mathbf{d}_{mn},\mathbf{\theta}\right)dh,
\end{equation}
These expressions are derived from the terms of the anisoplanatism error under
Rytov approximation\cite{2006JOSAA..23..613R}. The
weighting functions $W$ depend on SH geometry, statistical properties of the
turbulence, star separation $\mathbf{\theta}$, distance between subapertures
$\mathbf{d}_{mn}$ and altitude $h$. They can be seen as the response of the
system to a single layer at altitude $h$, for a certain distance between
subapertures and a certain star separation $\mathbf{\theta}$. Cross-correlations combine two directions
of analysis corresponding to the binary star separation $\mathbf{\theta}$, while auto-correlations correspond to
the case $\mathbf{\theta}=0$. Slope correlations are mainly sensitive to ground and low
altitude layers whereas scintillation correlations are more
sensitive to high altitude layers\cite{Vedrenne:07}, but there is no
scintillation on the pupil. Using
cross-correlations, the altitude resolution
$\delta h$ and the maximum sensing altitude $H_{max}$ are obtained with simple
geometrical rules\cite{2002MNRAS.337..103W} involving the subaperture
diameter $d_{sub}$ and telescope diameter $D$:
\begin{equation}
  \label{eq:resolution}
  \delta h\simeq \frac{d_{sub}}{\theta}
\end{equation}
\begin{equation}
  \label{eq:alt_max}
  H_{max}\simeq \frac{D}{\theta}
\end{equation}

Experimentally, correlations are estimated from a finite number of frames.
Then, they are arranged in a single dimension covariance vector
$\mathbf{C_{mes}}$, which is directly related to the $C_{n}^{2}$ profile according to
Eq. \ref{eq:direct_problem}:
\begin{equation}
  \label{eq:direct_problem}
  \mathbf{C_{mes}} = M \mathbf{C_n^2} + \mathbf{C_d} + \mathbf{u}
\end{equation}
where $M$ is the interaction matrix with column vectors formed by the concatenation of
the weighting functions $W$. $\mathbf{C_d}$ is the covariance vector of
detection noises affecting slope and intensity measurements. $\mathbf{u}$ is
the convergence noise representing uncertainties on $\mathbf{C_{mes}}$ due to the limited
number of frames. $\mathbf{C_{mes}}$ is estimated from measurements affected
by photon and detector noises that bias the correlation estimates. Assuming
the system is well calibrated, it is possible to completely
determine $\mathbf{C_d}$ and define a non-biased estimation of the correlation vector,
$\mathbf{\hat{C}_{mes}}=\mathbf{C_{mes}}-\mathbf{C_d}$.
Finally, this makes it possible to rewrite the
problem statement:
\begin{equation}
  \label{eq:new_direct_problem}
  \mathbf{\hat{C}_{mes}}= M\mathbf{C_{n}^{2}}+\mathbf{u}
\end{equation}
The covariance matrix of $\mathbf{u}$, $C_{conv}=\langle \mathbf{u}
\mathbf{u}^T\rangle$, is estimated from $\mathbf{C_{mes}}$ and $\mathbf{C_d}$
as the empirical covariance matrix of a Gaussian random variable vector. A
sampled estimate of $C_{n}^{2}$, $\mathbf{\tilde S}$, can be retrieved from
the inversion of Eq. \ref{eq:new_direct_problem}. Under positivity
constraint, because $C_{n}^{2}$ is never negative, $\mathbf{\tilde S}$
minimizes the maximum likelihood (ML) criterion $J$:
\begin{equation}
  \label{eq:ML_J} 
  J = (\mathbf{\hat{C}_{mes}} - M \mathbf{\tilde S})^T C_{conv}^{-1}
  (\mathbf{\hat{C}_{mes}} - M \mathbf{\tilde S})
\end{equation}
The diagonal of the covariance matrix $(M^TC_{conv}^{-1} M)^{-1}$ can be used as an
upper bound of the (square of the) sought error bars on the profile $\mathbf{\tilde S}$. The
implementation of these error bars will cover another material.

\section{Simulation of astronomical turbulent images}\label{sect:simu}

This section is dedicated to the description of an end-to-end simulation in
order to produce turbulent SH images of a binary star, taking into account
photon and detector noises. Parameters of the simulation are given in
Subsection \ref{sub:simu_para}. The numerical modeling is described in Subsection \ref{sub:modeling}.
Simulated images are finally presented in Subsection \ref{sub:sh_im} 

\subsection{Simulation parameters}\label{sub:simu_para}

\begin{figure}[!h]
  \begin{center}
    \resizebox{0.5\columnwidth}{!}{
      \includegraphics{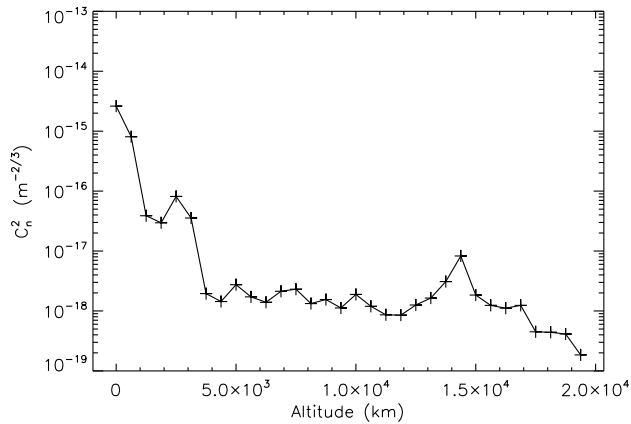}}
    \caption{Theoretical $C_n^2$ profile used for the simulation.}
    \label{fig:theoretical_profile}     
  \end{center}
\end{figure}

We consider a $D=1.5$ meter telescope with $30~\%$ of central obscuration, and two SH geometries,
$30 \times 30$ and $15 \times 15$ subapertures. The subaperture diameter is $d_{sub}=5~cm$ or
$d_{sub}=10~cm$, depending on the number of subapertures. The wavelength is
$\lambda=0.55~\mu m$. The observed object is a binary star with separation
$\theta=20~''$ modeled by a
two-point source. We assume a difference of one magnitude between the two
stars. Fluxes are about $120$ and $300$ photons per subaperture and per frame
for each star. The $C_n^2$ profile is typical of an astronomical site, with strong
turbulence at ground level, and turbulent activity in altitude between $13$
and $15~km$ (Fig. \ref{fig:theoretical_profile}). It is composed of $32$
$C_{n}^{2}$ values defined at $32$ altitudes (\textit{i.e.} 32 layers). The resultant Fried
parameter $r_{0}\simeq 5~cm$. Outer scale $L_{0}=8~m$
and inner scale $l_0=5~mm$. Simulation outputs are $660 \times 660$
Shannon sampled images, leading respectively to $22 \times 22$ and $44 \times
44$ pixels per subaperture, for each SH geometry.

\subsection{Numerical modeling}\label{sub:modeling}

The numerical modeling is composed of two parts, the first consists of the
diffractive propagation and the second contains the image formation
process. Here we expose briefly how it works, more details can be found in
Ref.\cite{2006JOSAA..23..613R}.
In order to simulate the propagation through turbulence, we use
the PILOT (Propagation and Imaging, Laser and Optics through Turbulence) code,
developed at Onera. The source is sampled on a bidimensionnal Cartesian grid
of points. The turbulent volume is broken down into a series of discrete
layers. The code simulates turbulent phase screens, representing turbulent layers,
following the von Karman spectrum of the refractive-index fluctuations, and
Fresnel propagation takes place between the screens. Therefore we consider
both phase and scintillation effects. The output of PILOT is a
complex electromagnetic field at the telescope pupil. It is then sampled at
SH subaperture level. Point-spread functions (PSF) for each point of the source are then
obtained by computing the square modulus of the Fourier transform of the
electromagnetic field. The final turbulent image in a SH subaperture is
built with the sum of the PSFs.

\subsection{Shack-Hartmann images}\label{sub:sh_im}

\begin{figure}[!h]
  \begin{center}
    \resizebox{0.5\columnwidth}{!}{
      \includegraphics{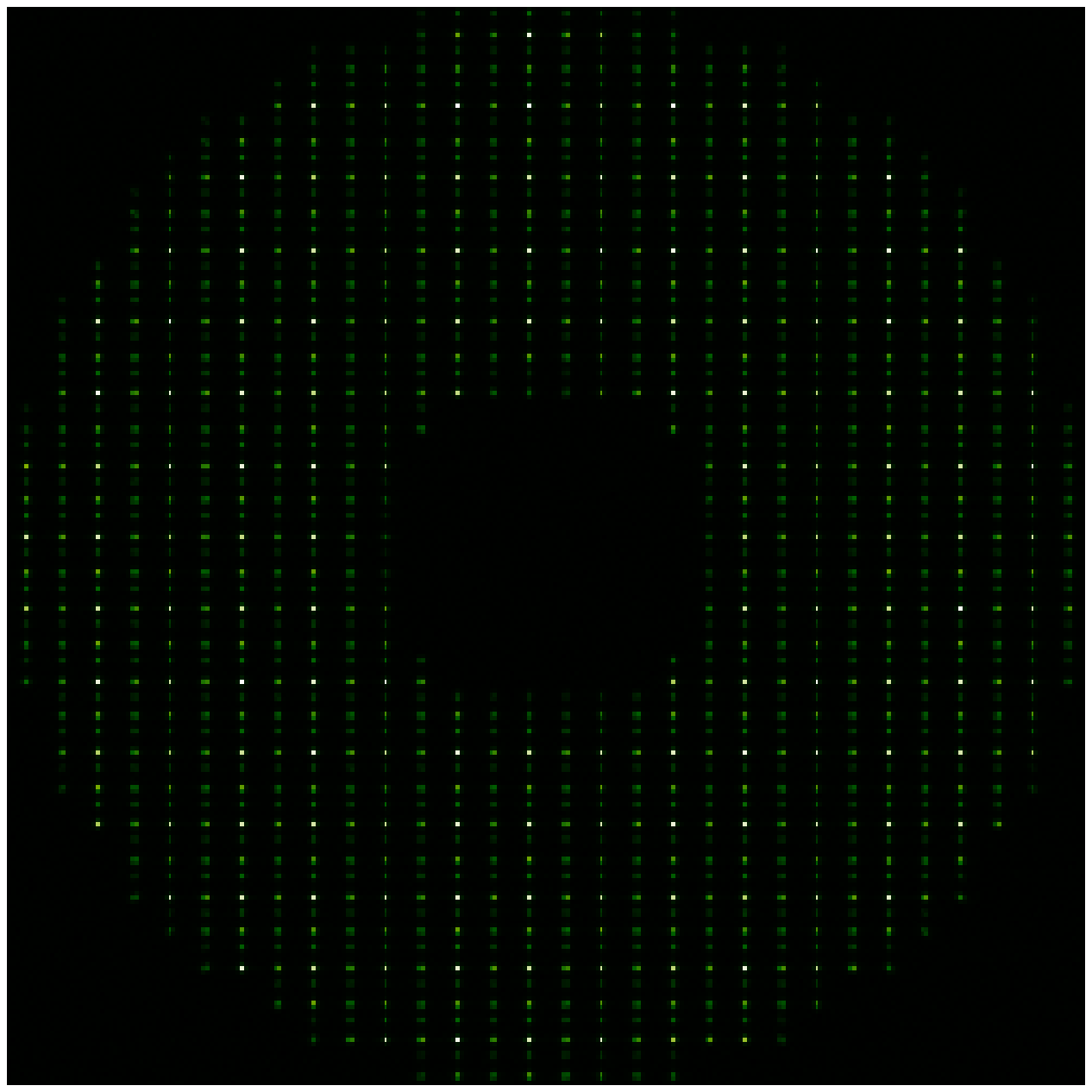}
      \includegraphics{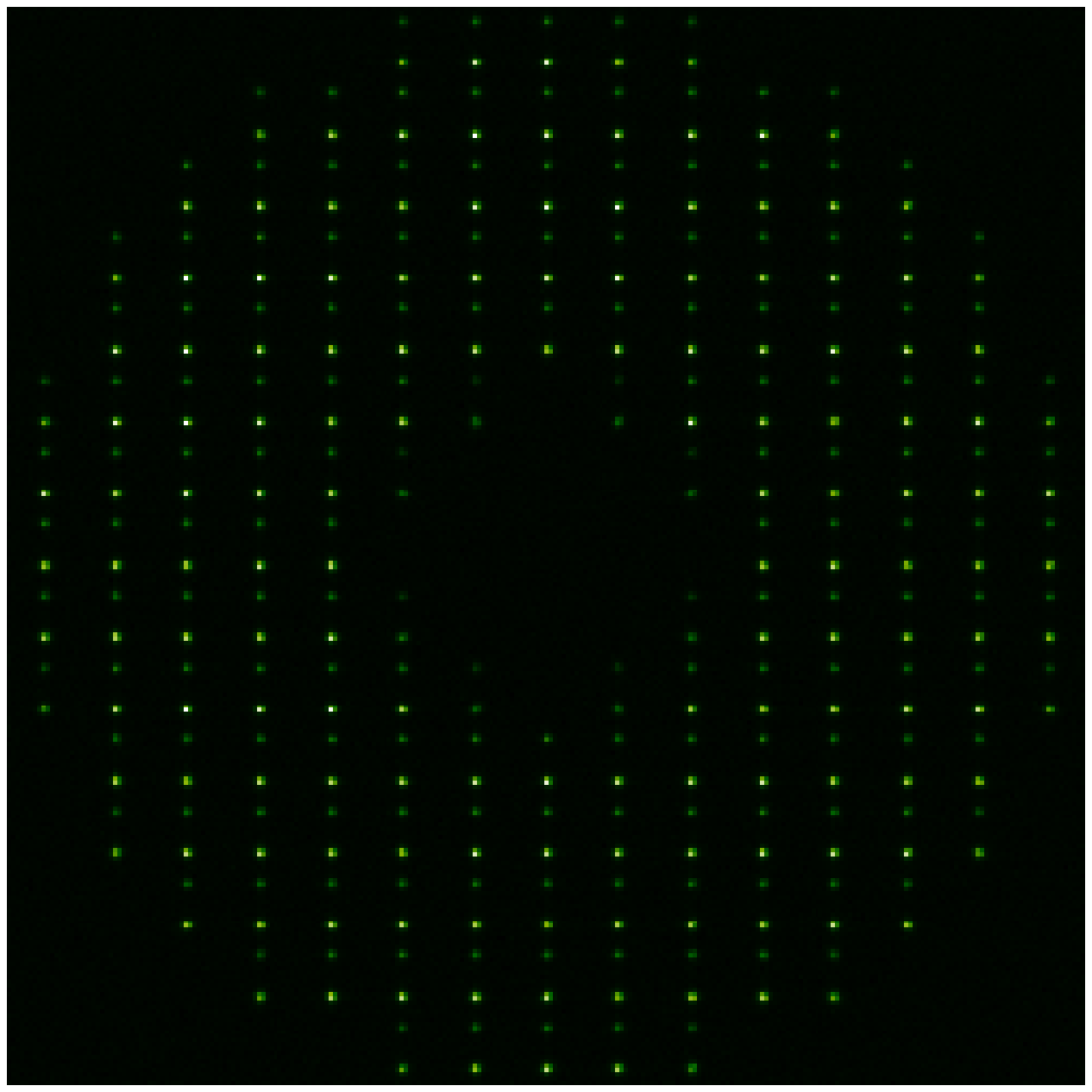}}
    \resizebox{0.5\columnwidth}{!}{      
      \includegraphics{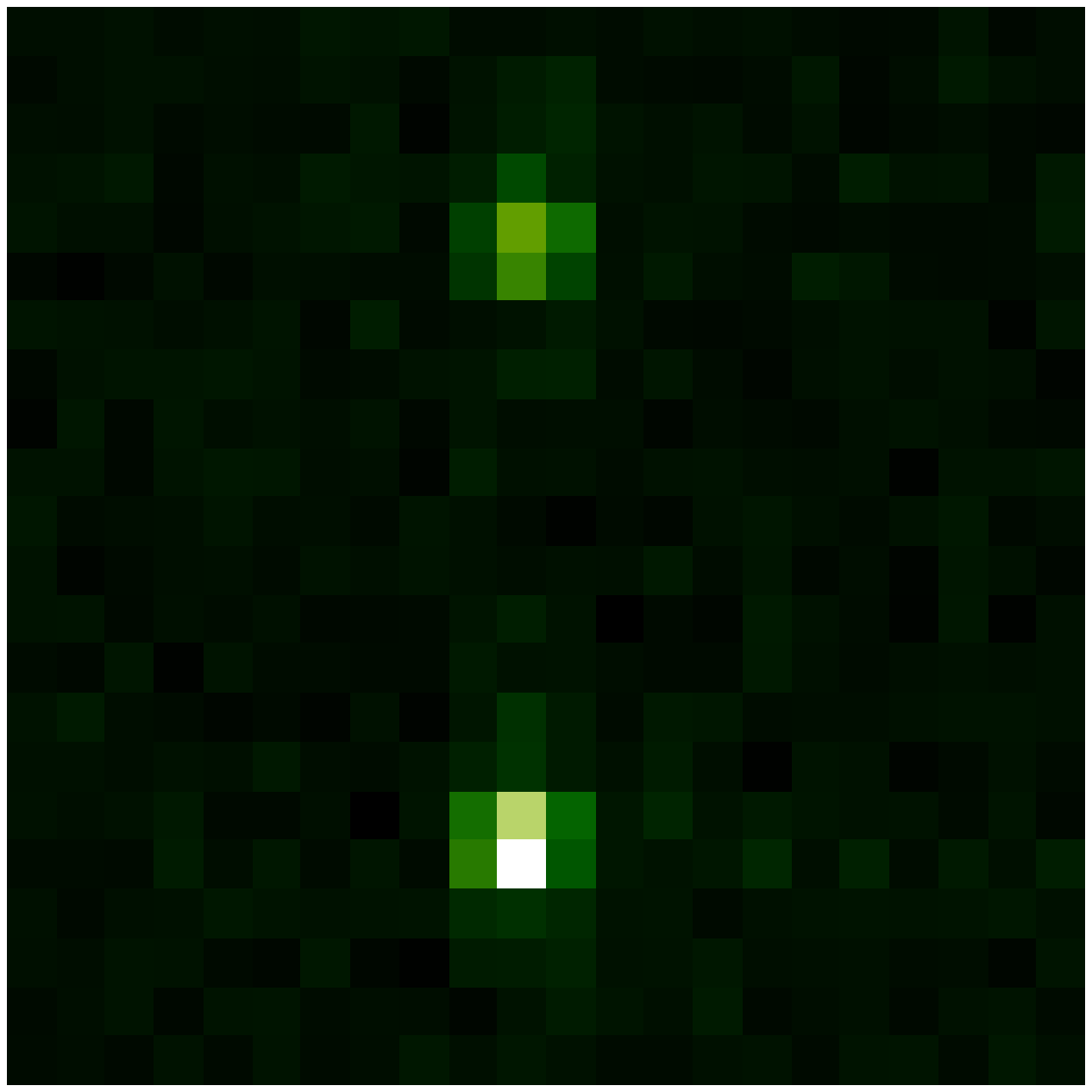}
      \includegraphics{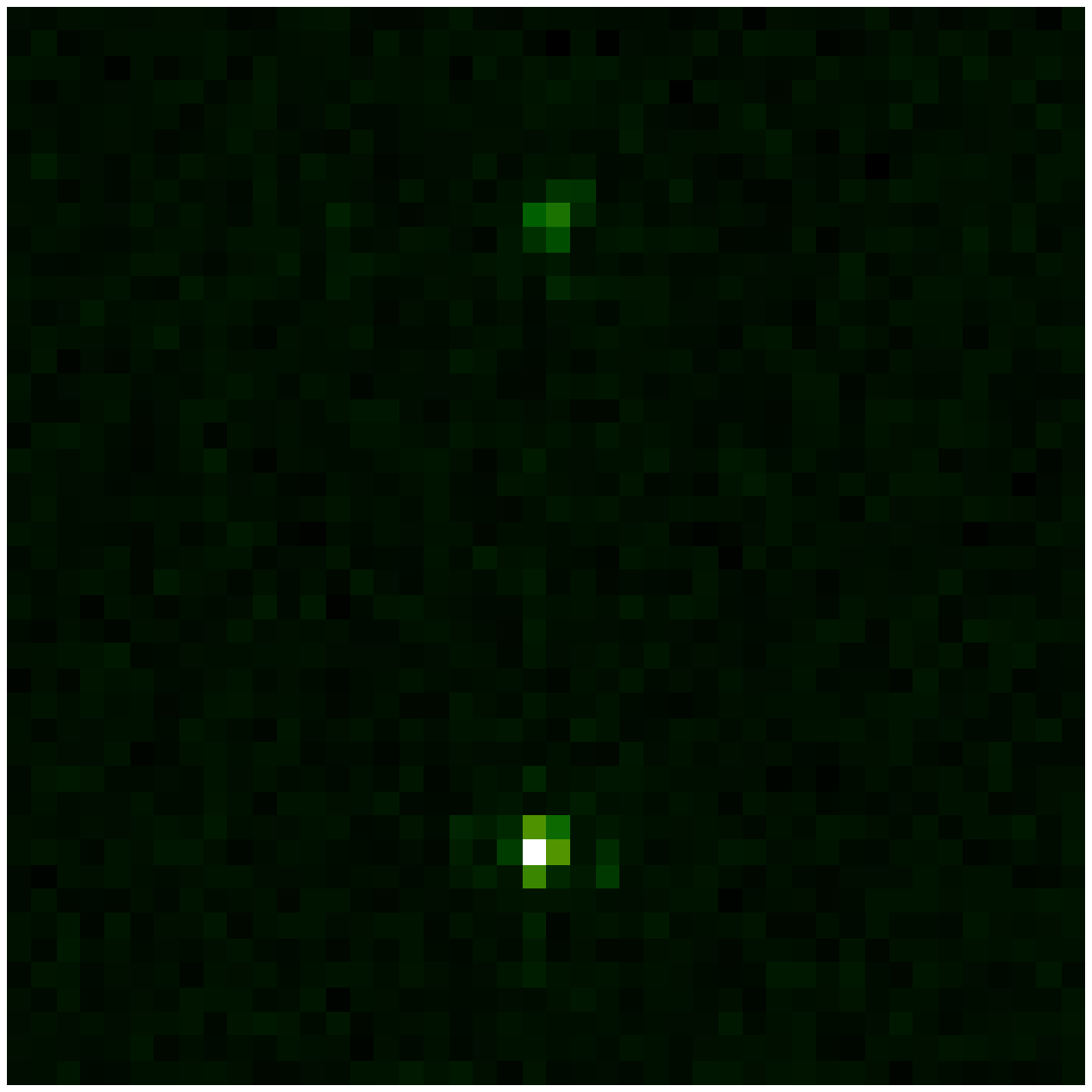}}
    \caption{Full SH images (averaged over 100 frames) and short-exposure
      subaperture images, obtained after propagation
      through turbulence and addition of detection noises. Left: $30
      \times 30$ geometry. Right: $15\times 15$ geometry.}
    \label{fig:sh_im}     
  \end{center}
\end{figure}

$30 \times 30$ and $15 \times 15$ SH images are computed after $100$ wave propagations
through $32$ phase screens, representing the $32$ turbulent layers of the $C_n^{2}$
profile in Fig. \ref{fig:theoretical_profile}. Then, $9$ cuts are performed on the
electromagnetic field after each propagation, in order to get $900$ SH frames.
These images are noise-free. Detection noises are added outside of the
simulation code. Photon noise is modeled by a Poisson law and
detector noise, assumed to be $\sigma_{e^-}=1~e^-$ per pixel and per frame, is modeled by
a Gaussian law. Here, we assume that noise is statistically
independent for different subapertures and stars. Typical SH frames and subaperture images are
presented in Fig. \ref{fig:sh_im}.  

\section{Data processing}\label{sect:data_processing}

In this section we describe how we measure slopes and scintillation indexes
from simulated SH images in Subsection$~$\ref{sub:data_extrac}, in order to make
correlation maps. The bias subtraction process is detailed in Subsection$~$\ref{sub:bias_subtraction}.

\subsection{Slopes and scintillation indexes extraction}\label{sub:data_extrac}

Wavefront slopes and scintillation indexes for each star are calculated in each subaperture for all SH
short-exposure images. Subapertures that are less than $95$ per cent
illuminated are excluded from the analysis. To compute slopes
$\mathbf{s_{m}}\left(\mathbf{\alpha}\right)$ in each of the two orthogonal ``tip''
and ``tilt'' directions, we use a centre of gravity (COG) algorithm. In order
to separate the contributions of the two stars, and to limit the effects of noise, centroids are evaluated by delimiting windows around the stars. Boxes
of $4 \times 4$, $6 \times 6$ and $8 \times 8$ pixels are considered for the
$30\times 30$ geometry, where the spots are not so much distorted because of
turbulence ($\frac{d_{sub}}{r_{0}} \sim 1$). For the $15 \times 15$ geometry, spots are more distorted. Indeed,
the subaperture diameter is two times larger ($\frac{d_{sub}}{r_{0}} \sim 2$), and images are more affected by
turbulence. So, we have to chose larger windows, and size of $8 \times 8$, $10
\times 10$, $12 \times 12$, $14 \times 14$ and $16 \times 16$ pixels are considered.
The positions of the boxes are determined by locating the mean position of the maximum of the
spots in the subapertures. The total intensity corresponds to the total of all
pixel intensities into the box. Scintillation index $\delta i_{m}\left(\mathbf{\alpha} \right)$ is derived by subtracting the mean
intensity in the subaperture over the whole sequence of images
$o_{m}\left(\mathbf{\alpha} \right)$, and by dividing
the result by this same term. Slopes and
scintillation indexes are extracted from noise-free and noisy images for
comparison. Noise propagation occurred in the computation of slopes and
scintillation indexes. As they are measured using a COG algorithm, noises
contribution are then given by\cite{1987LFTR...28...17R}:
\begin{equation}
  \label{eq:slonoise_phot} 
  \sigma_{\Delta \phi_{phot}}^{2} = \frac{\pi^{2}}{2}\frac{1}{N_{ph}}\left(\frac{N_{T}}{N_{D}}\right)^{2}
\end{equation}
\begin{equation}
  \label{eq:slonoise_det} 
  \sigma_{\Delta \phi_{det}}^{2} = \frac{\pi^{2}}{3}\left(\frac{\sigma_{e^{-}}}{N_{ph}}\right)^{2}\left(\frac{N_{s}^{2}}{N_{D}}\right)^{2}
\end{equation}
for slope measurements. $N_{T}$ is the image full width at half maximum (FWHM), $N_{D}$ is the FWHM limited
by diffraction and $N_{S}$ is the number of pixels for the COG
calculation. However, in Eq. \ref{eq:slonoise_phot}, $\sigma_{\Delta \phi_{phot}}^{2}$ does not depend on the
size of the window. It has been shown that when
$N_{S} \ge 2 N_{D}$, $\sigma_{\Delta \phi_{phot}}^{2}$ can be written as\cite{2006MNRAS.371..323T}:
\begin{equation}
  \label{eq:slonoise_phot_corr} 
  \sigma_{\Delta \phi_{phot}}^{2} = \frac{2}{N_{ph}}\left(\frac{N_{S}}{N_{D}}\right)
\end{equation}
For scintillation measurements we have:  
\begin{equation}
  \label{eq:scinoise_phot} 
  \sigma_{\delta i_{phot}}^{2} = \frac{1}{N_{ph}}
\end{equation}
\begin{equation}
  \label{eq:scinoise_det} 
  \sigma_{\delta i_{det}}^{2} = N_{S}^{2}\left(\frac{\sigma_{e^{-}}}{N_{ph}}\right)^{2}
\end{equation} 
For each kind of data, the total noise, $\sigma_{\Delta \phi_{noise}}^{2}$ for
slopes and $\sigma_{\delta i_{noise}}^{2}$ for scintillation, is the sum of the terms
$\sigma_{phot}^{2}$ and $\sigma_{det}^{2}$.
The effects of the size of the window and the effects of noise on the $C_{n}^{2}$ measurement
are investigated in Subsections \ref{sub:win_influence} and \ref{sub:noise_influence}.

\subsection{Making of correlation maps and bias subtraction process}\label{sub:bias_subtraction}

\begin{figure}[!h]
  \begin{center}
    \resizebox{0.5\columnwidth}{!}{
      \includegraphics{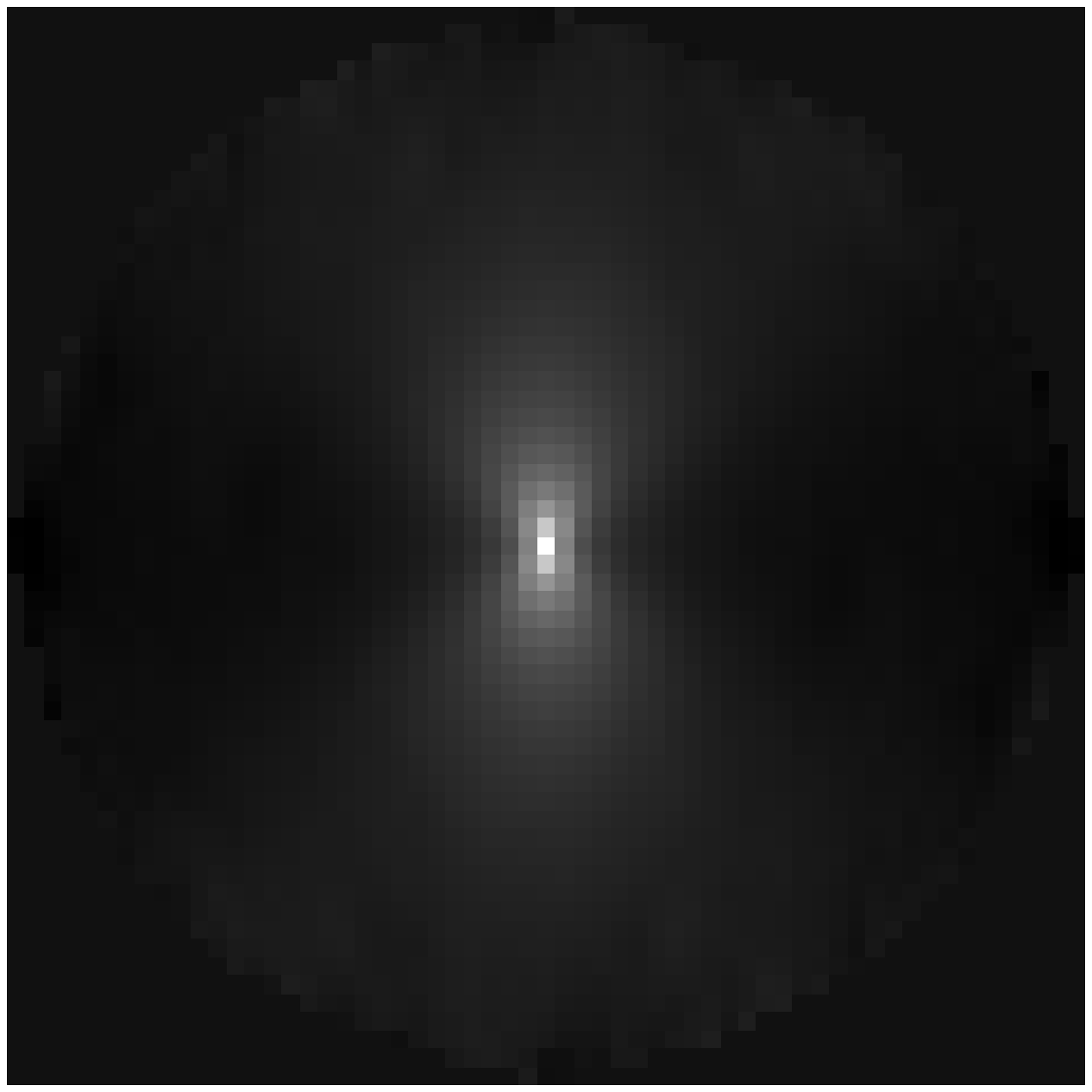}
      \includegraphics{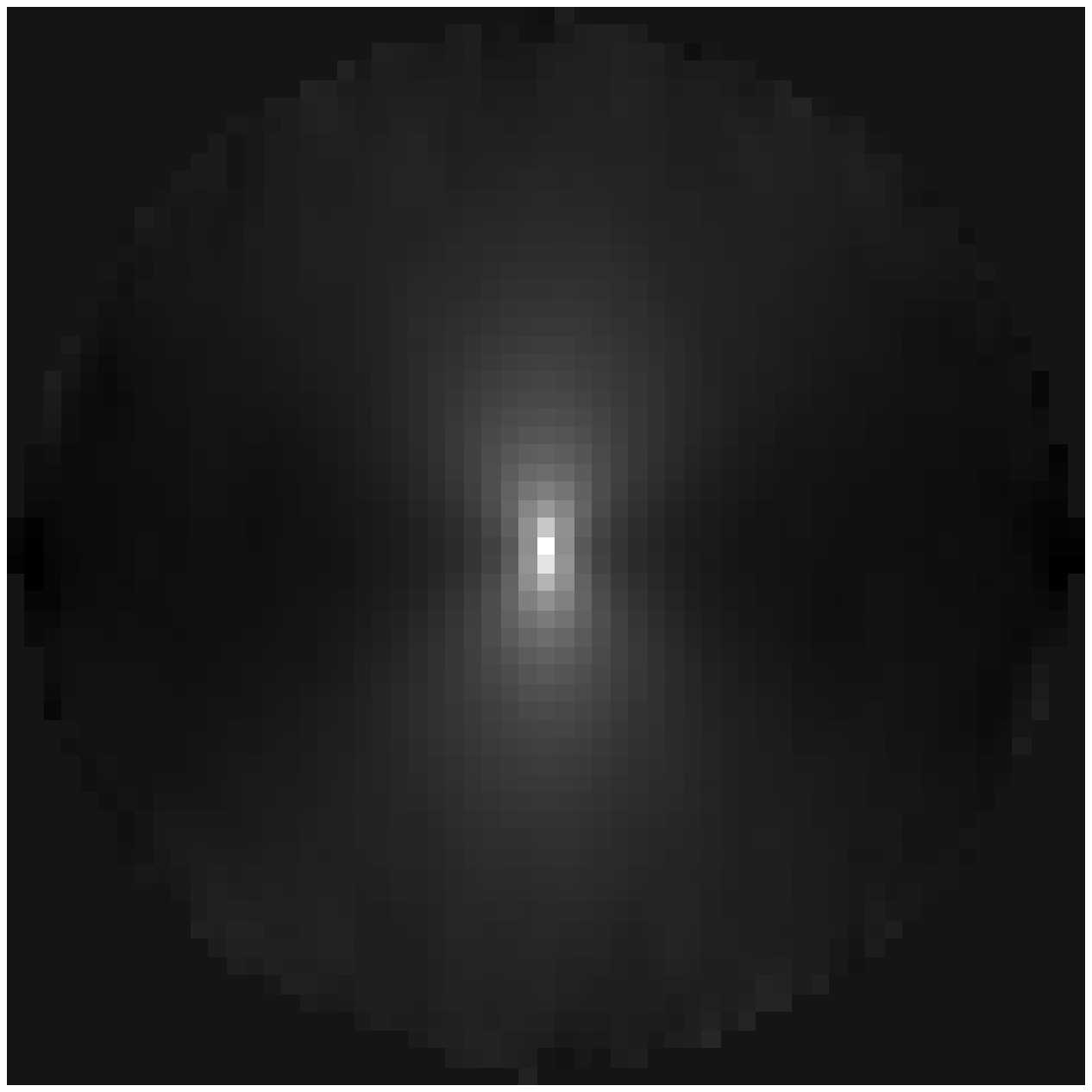}}
    \resizebox{0.5\columnwidth}{!}{      
      \includegraphics{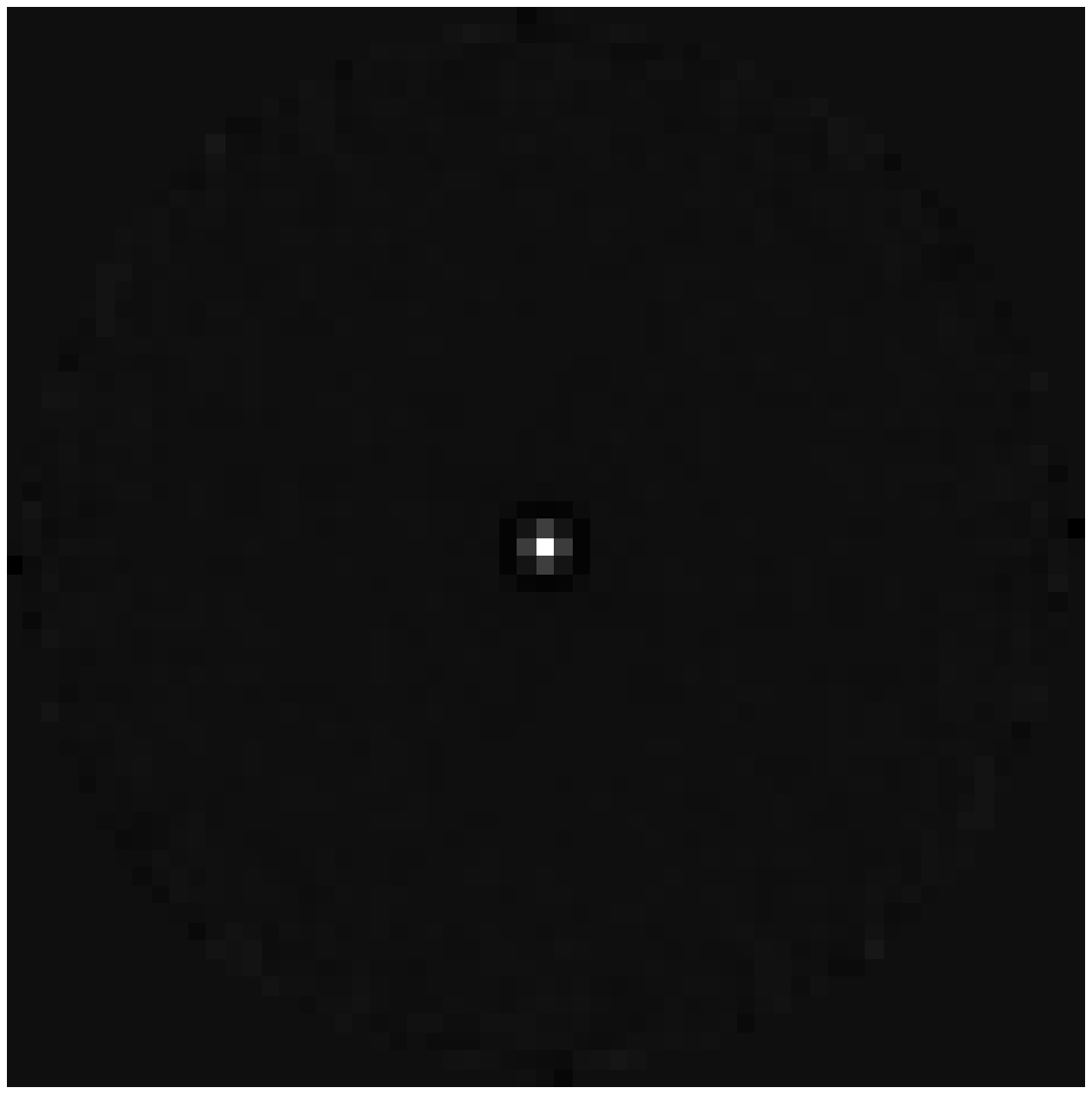}
      \includegraphics{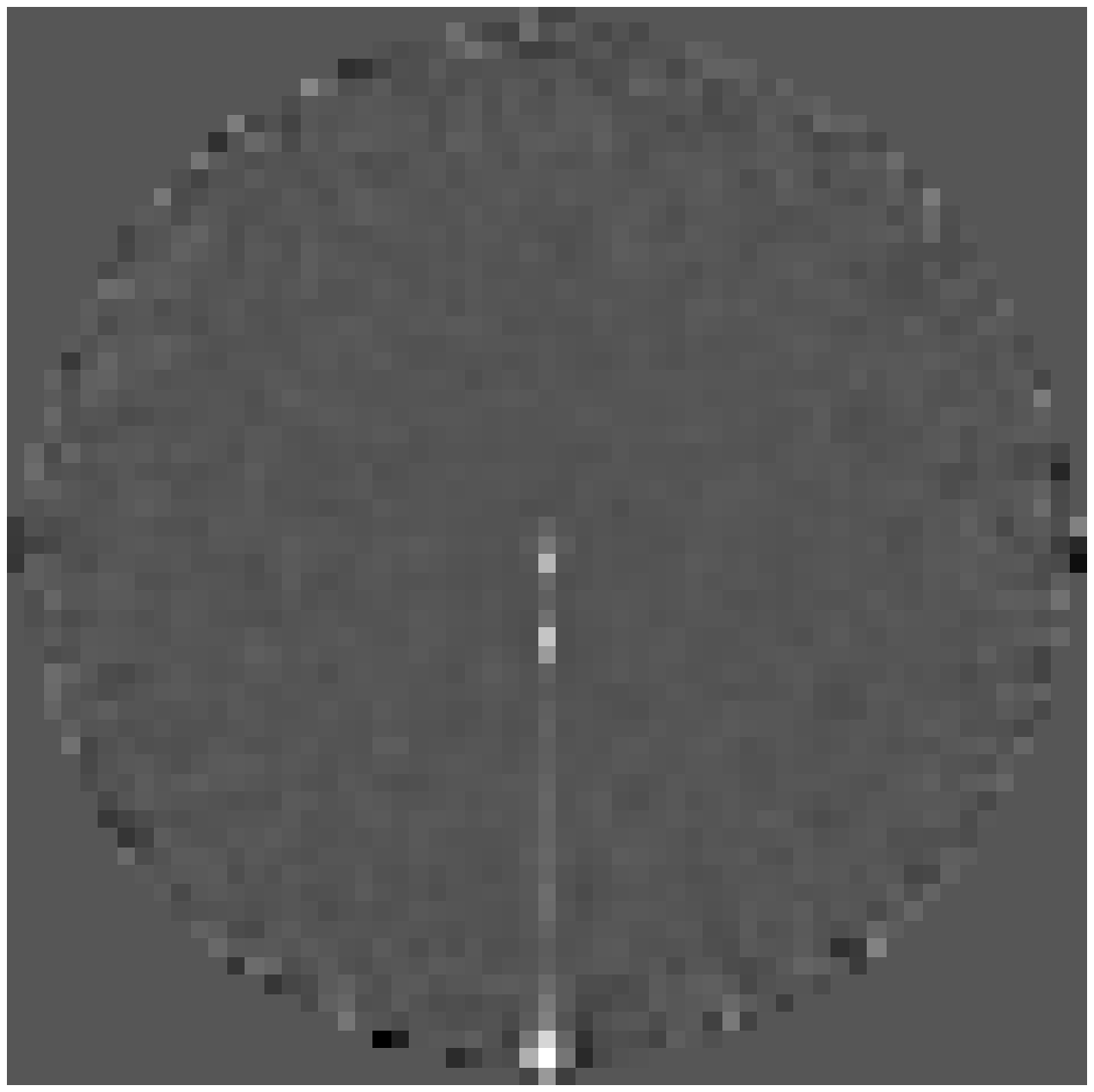}}
    \caption{Correlation maps for the $30 \times 30$ geometry. The maps
      dimensions are 2 $\times$ number of subapertures-1, illustrating
      the correlation for all couples of subapertures. Top-left: slope
      auto-correlations in the x direction. Top-right: slope
      cross-correlations in the x direction. Bottom-left: scintillation
      auto-correlations. Top-right: scintillation cross-correlations.}
    \label{fig:corr_map}     
  \end{center}
\end{figure}

Auto-correlations (on a single star) and cross-correlations (between the two stars) are computed from slopes and
scintillation indexes. In this simulated case, auto-correlations are evaluated for
each star, and averaged, which increases the statistics by doubling the number
of samples. Correlations are calculated for all separations between
subapertures and represented as correlation maps. These maps show the
mean correlation between all couples of subapertures having the same gap
(Fig. \ref{fig:corr_map}), for the two kind of data, slopes and
scintillation indexes. In the presence of measurement noises, these maps are biased.
Assuming that the noise is statistically independent for different stars and subapertures, only the
variance, namely the central value of the auto-correlation maps of slopes and scintillation are
biased. A good way to estimate the bias
is then to analytically calculate the noise using Eqs. \ref{eq:slonoise_phot},
\ref{eq:slonoise_det}, \ref{eq:slonoise_phot_corr}, for slope
auto-correlations and Eqs. \ref{eq:scinoise_phot},
\ref{eq:scinoise_det}, for scintillation auto-correlations, knowing the number of photons $N_{ph}$ and the detector
noise $\sigma_{e^{-}}$. As the two stars have different magnitudes, we apply different
numbers of photons per subaperture and per frame in the formulae. The final bias is
the mean of the biases given by each star. Once it is calculated, we just
have to subtract it from the central value of the auto-correlation maps. The
effect of the bias subtraction on the $C_{n}^{2}$ retrieval will be studied in
Subsection \ref{sub:noise_influence}.

\section{$C_n^2$ restoration: results}\label{sect:results}

The $C_{n}^{2}$ profiles reconstructed with the two SH geometries are
presented and discussed in this section. We begin by studying the influence of
the size of the window in Subsection \ref{sub:win_influence}. Then, the effects of
measurement noises and bias subtraction are analyzed in Subsection
\ref{sub:noise_influence}. We finally compare the results with those obtained
with other methods in Subsection \ref{sub:comparisons}, and the advantages of
the two SH geometries are examined.

\subsection{Influence of the COG window}\label{sub:win_influence}

\begin{figure}[!h]
  \begin{center}
    \resizebox{0.9\columnwidth}{!}{
      \includegraphics{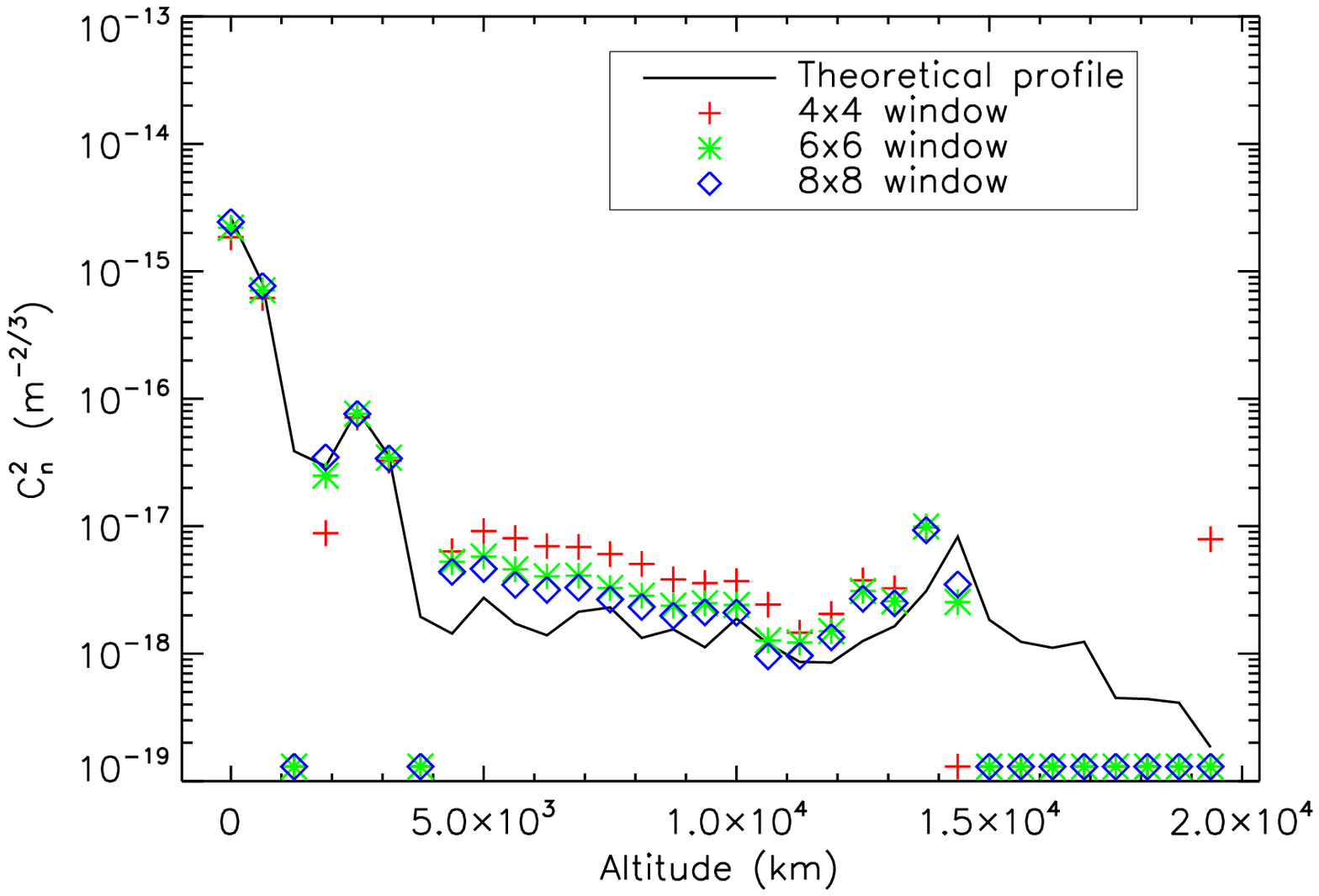}
      \includegraphics{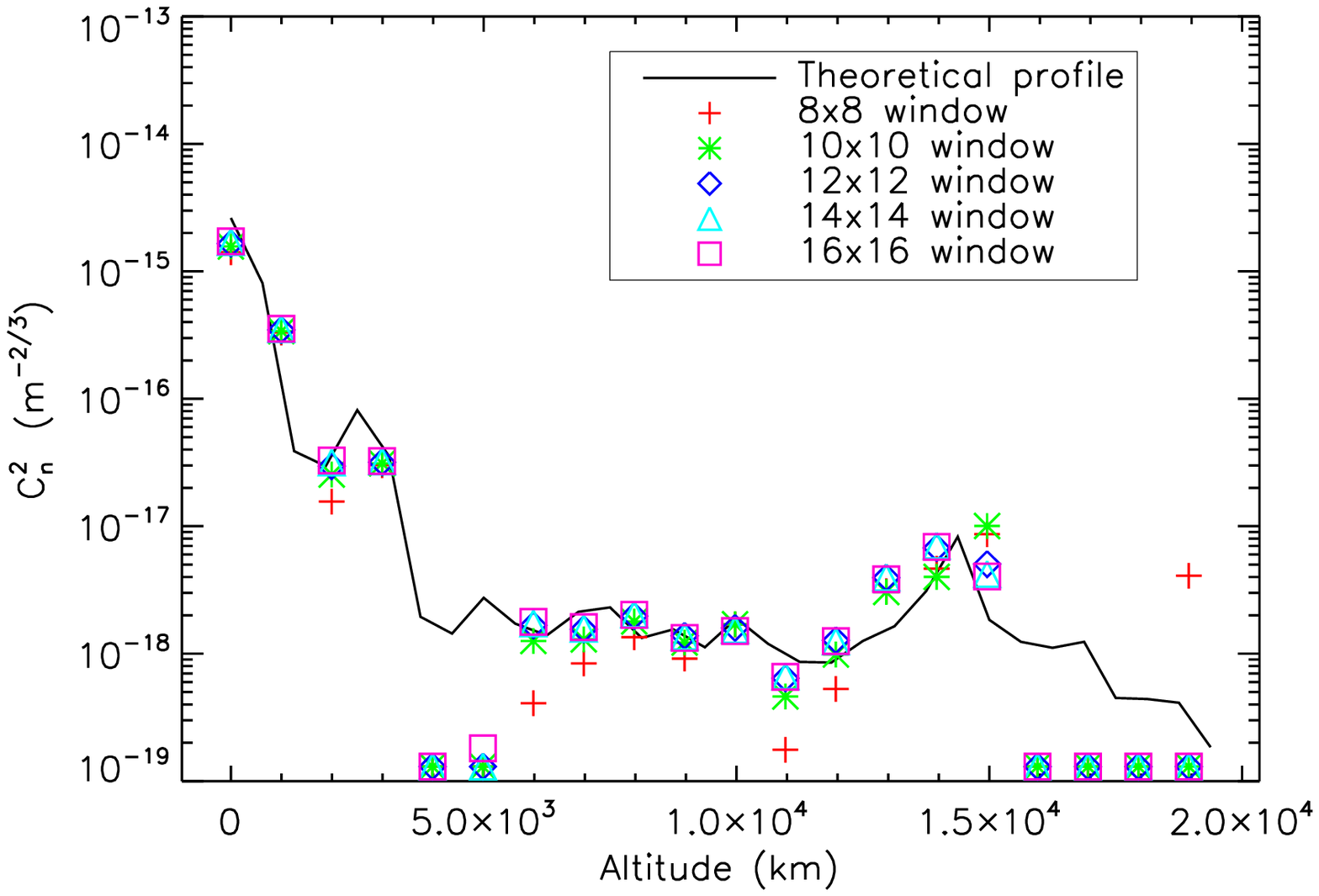}}
    \caption{Impact of the size of the window for COG and intensity
      calculation on the reconstructed $C_{n}^{2}$ profile. Left: $30 \times
      30$ geometry. Right: $15 \times 15$ geometry.}
    \label{fig:impact_fen}     
  \end{center}
\end{figure}

As we said in Subsection \ref{sub:data_extrac}, windows of different sizes are
considered to compute slopes and intensities. Here, we only consider noise-free data. $C_{n}^{2}$ profiles reconstructed with the two
geometries, for different sizes of window, are shown in Fig.
\ref{fig:impact_fen}. $C_{n}^{2}$ values lower than $1\times
10^{-19}~m^{-\frac{2}{3}}$ are automatically put to $1.3\times
10^{-19}~m^{-\frac{2}{3}}$ on the graph, for better understanding of the estimation. With the
$30\times 30$ geometry we restore $32$ layers, that is to say the nominal number of layers
of the input theoretical $C_{n}^{2}$ profile, while with the $15\times 15$
geometry, we only restore $20$ layers.  
We can notice immediately that the size of the window introduces a slight bias
in the $C_{n}^{2}$ profile measurement, and that for the two
SH geometries. For the $30 \times 30$ geometry, although the low altitude layers are
almost well reconstructed with all sizes of window,
the greatest differences occurred after $4~km$ of altitude. $C_{n}^{2}$ values
are always over-estimated, but the smaller is the box, the higher is the
over-estimation. The layer at $14~km$ is imperfectly
reconstructed with the $6\times 6$ and $8\times 8$ windows, while it is not at
all with the $4\times 4$ window. Layers after $15~km$ of altitude are not well
estimated with the three sizes of window (\textit{cf.} Subsection \ref{sub:comparisons}). For the $15 \times 15$ geometry, we
also observe that the reconstruction is improved when the size of the box
increases. Identically, low altitude layers are almost well reconstructed. But,
the plateau between $6~km$ and $12~km$ is better estimated when the box dimensions
reach $10\times 10$ pixels. Results are quite similar for all sizes of window
for layers between $12~km$ and $14~km$. The layer at $15~km$ is better evaluated
when the box size reaches $12\times 12$ pixels. As for the other geometry, layers after $15~km$ of altitude are not well
estimated with all sizes of window. Windows of $4\times 4$ and $6\times 6$
pixels were also tested, but results were too much biased so they are not
presented.

This bias depending on the size of the window can be explained. When
calculating slopes and intensities in little boxes, smaller than the whole
subaperture, the spot is partially truncated. So, part of the flux is not taken into
account for the COG calculation and the total intensity evaluation, leading to
a biased estimation, that propagates in the $C_{n}^{2}$ profile retrieval.
As a consequence, the size of the window must be chosen carefully. It must
depend on the spot distortion and the signal to noise ratio (SNR) in the subaperture.
The more the spot is distorted, the more the box has to be large, in order not
to truncate too much the spot for the measurements. However, a larger
window introduces more noise in the measurements (Eqs.
\ref{eq:slonoise_det}, \ref{eq:slonoise_phot_corr}, \ref{eq:scinoise_det}).
Moreover, if stars are too closed, the size of the box
is implicitly limited. In the following, we keep windows of $8\times 8$
pixels for the $30\times 30$ geometry and windows of $12\times 12$ pixels for
the $15\times 15$ geometry.

\subsection{Influence of noise and effect of bias subtraction}\label{sub:noise_influence}

\begin{figure}[!h]
  \begin{center}
    \resizebox{0.9\columnwidth}{!}{
      \includegraphics{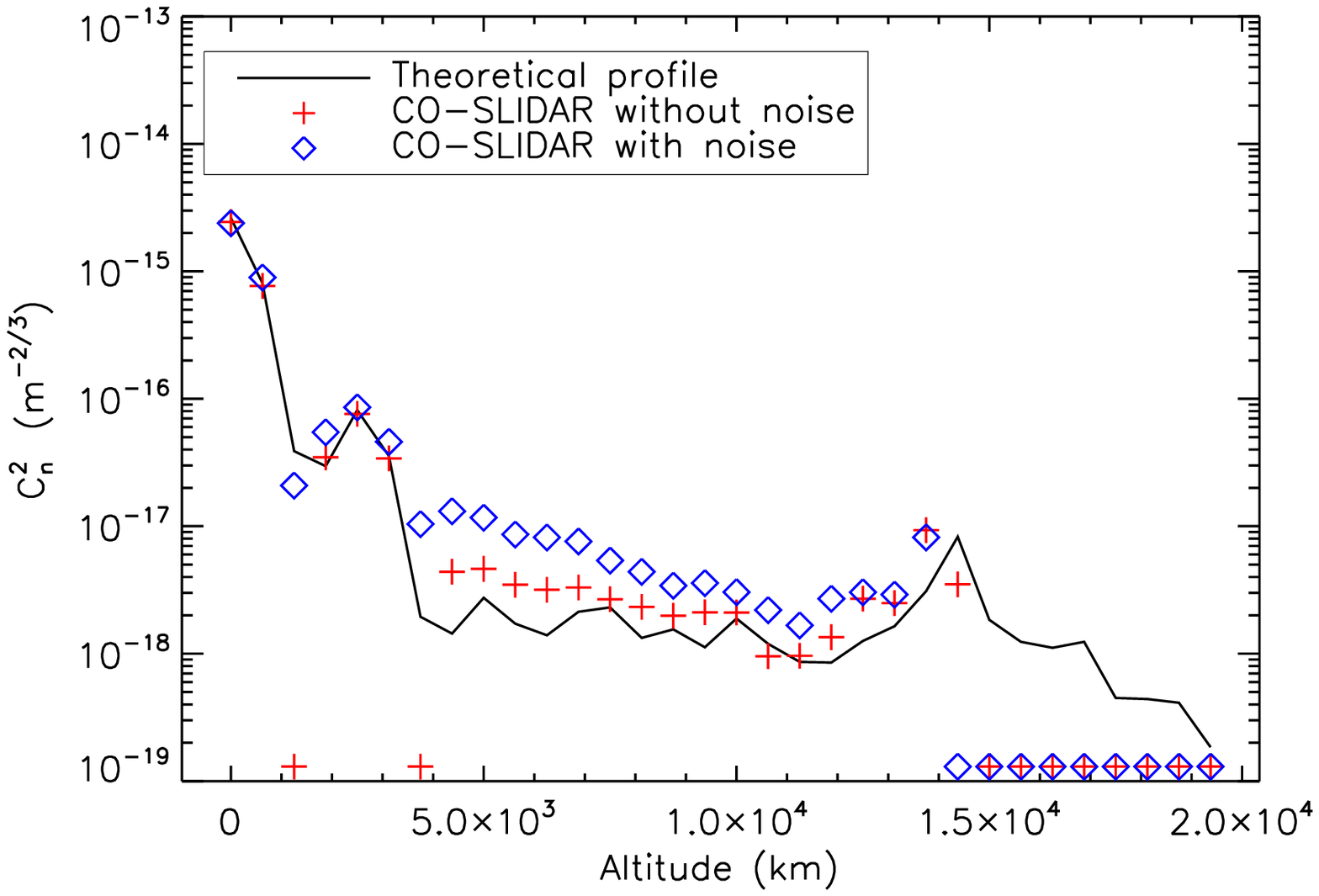}
      \includegraphics{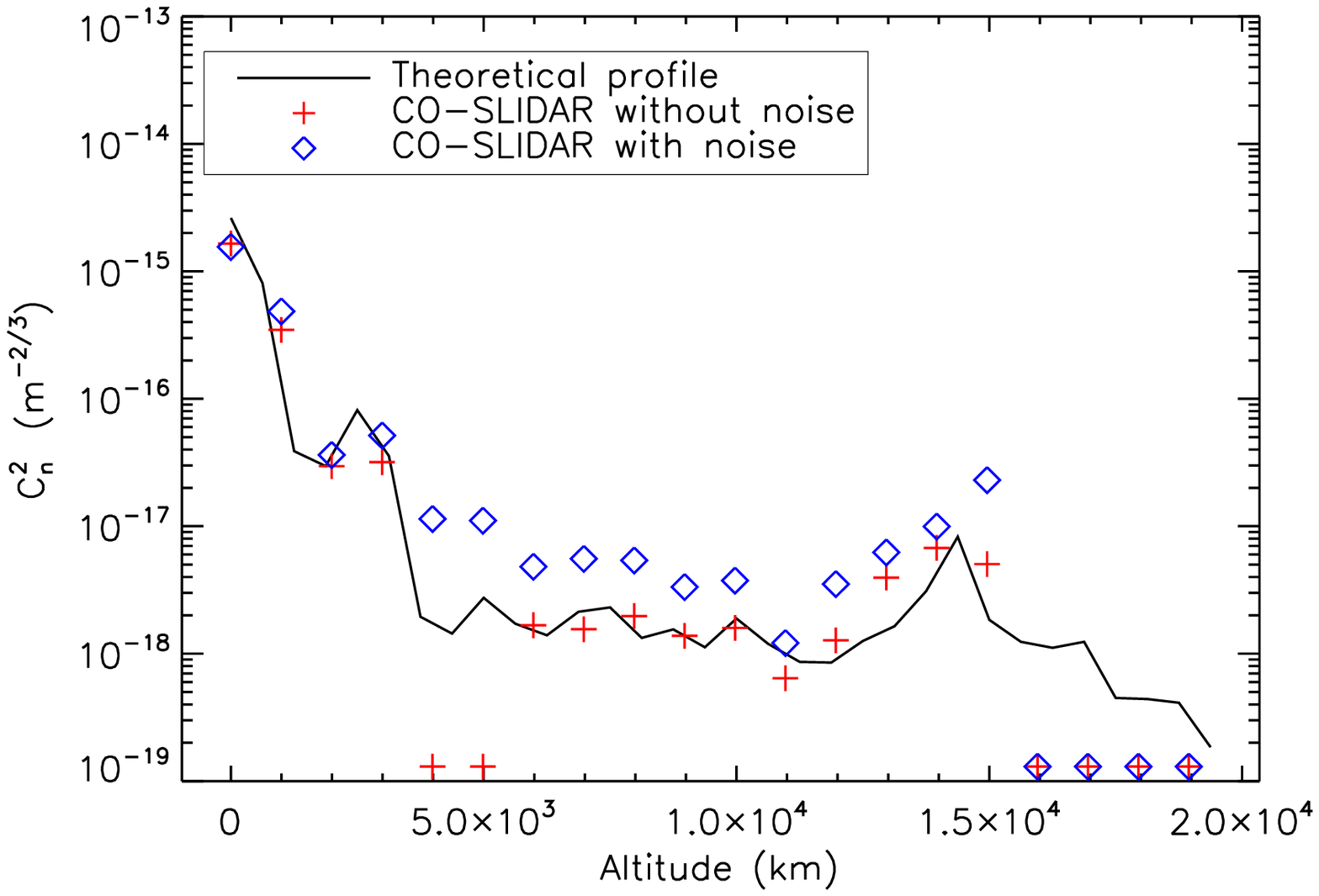}}
    \caption{Influence of noise on the $C_{n}^{2}$ profile reconstruction. Left: $30 \times
      30$ geometry. Right: $15 \times 15$ geometry.}
    \label{fig:impact_bruit}     
  \end{center}
\end{figure}

Now we perform $C_{n}^{2}$ reconstruction from noisy data. The impact of noise
propagation on the results is presented in Fig. \ref{fig:impact_bruit}.
Noise adds an offset to all estimated $C_{n}^{2}$ values, with both
geometries. Low altitude layers, until $4~km$ of altitude, are less impacted
by this offset than layers between $4$ and $15~km$. Actually, these layers are
mostly restored thanks to the scintillation signal. But, this signal has a
poorer SNR than the slope signal. For the two geometries, the SNR on slopes
and scintillation signals are detailed in Tables \ref{tab:SNR_slo} and
\ref{tab:SNR_sci}, using Eqs. \ref{eq:slonoise_det},
\ref{eq:slonoise_phot_corr}, \ref{eq:scinoise_phot}, \ref{eq:scinoise_det}). These
tables show that there is a good SNR on slope signal $\sigma_{\Delta \phi_{turb}}^{2}$, but it is very weak on
the scintillation signal $\sigma_{\delta i_{turb}}^{2}$, even smaller than $1$ in the case of the $15\times
15$ geometry. This observation explains the bad estimation of the $C_{n}^{2}$ profile
at altitudes between $4$ and $15~km$.
\begin{table}[!h]
  \begin{center}
    \begin{tabular}{c|c|c|}
      \cline{2-3}
      & SH $30\times 30$   &  SH $15\times 15$  \\
      \hline
      \multicolumn{1}{|c|}{$\sigma_{\Delta \phi_{turb}}^{2}$ ($rad^{2}$)} &
      $3.4$ & $9.5$  \\
      \hline
      \multicolumn{1}{|c|}{$\sigma_{\Delta \phi_{noise}}^{2}$ ($rad^{2}$)} &
      $0.2$ & $0.8$ \\
      \hline
      \multicolumn{1}{|c|}{$SNR$} & $17$ & $12$ \\
      \hline
    \end{tabular}
    \caption{SNR on slope signal, for the two SH geometries.}
    \label{tab:SNR_slo} 
  \end{center}
\end{table}

\begin{table}[!h]
  \begin{center}
    \begin{tabular}{c|c|c|}
      \cline{2-3}
      & SH $30\times 30$   &  SH $15\times 15$  \\
      \hline
      \multicolumn{1}{|c|}{$\sigma_{\delta i_{turb}}^{2}$} &
      $2.2 \times 10^{-2}$ & $8.3\times 10^{-3}$  \\
      \hline
      \multicolumn{1}{|c|}{$\sigma_{\delta i_{noise}}^{2}$} &
      $8.7\times 10^{-3}$ & $1.2 \times 10^{-2}$ \\
      \hline
      \multicolumn{1}{|c|}{$SNR$} & $2.7$ & $0.7$ \\
      \hline
    \end{tabular}
    \caption{SNR on scintillation signal, for the two SH geometries.}
    \label{tab:SNR_sci} 
  \end{center}
\end{table}

\begin{figure}[!h]
  \begin{center}
    \resizebox{0.9\columnwidth}{!}{      
      \includegraphics{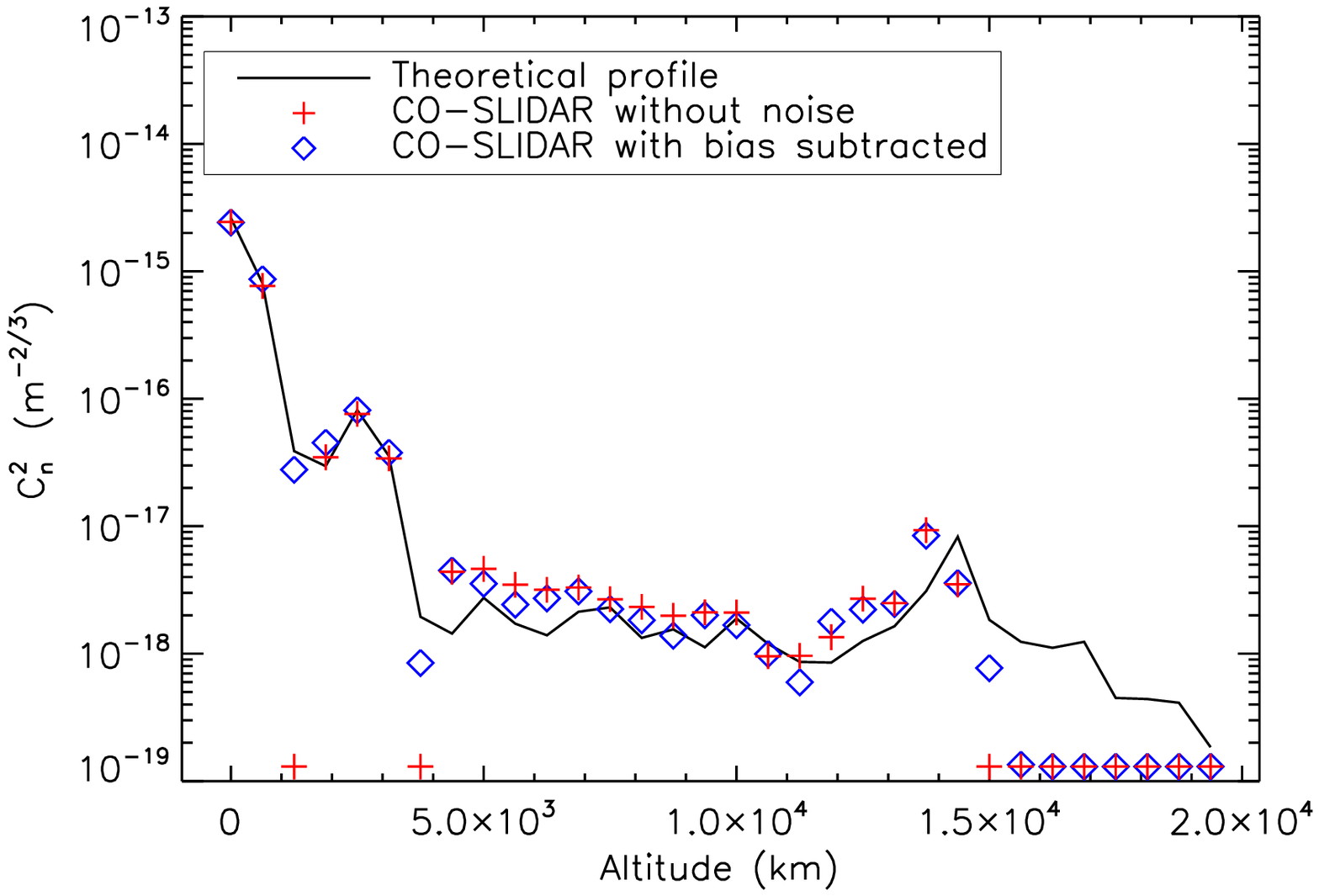}
      \includegraphics{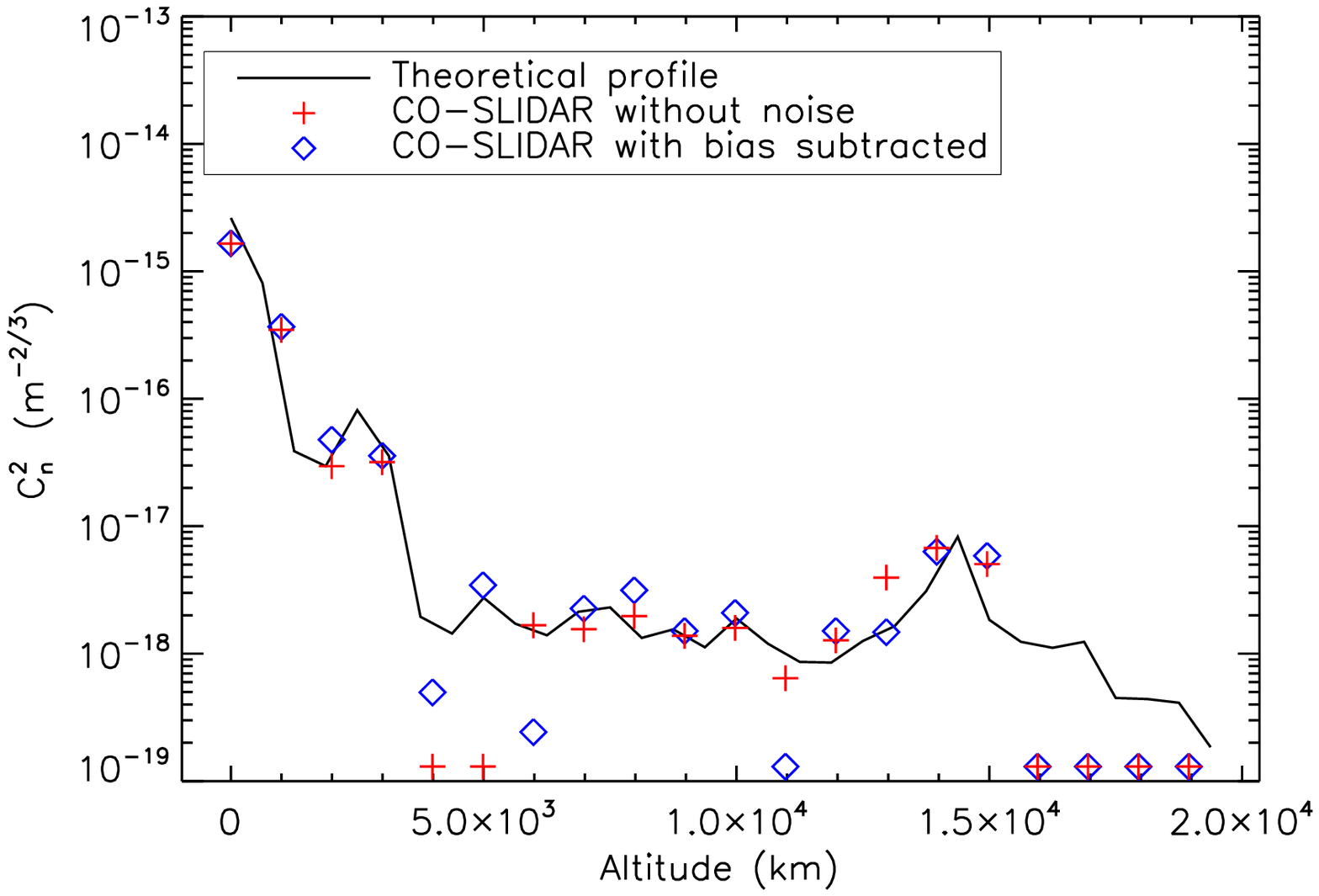}}
    \caption{Impact of bias-subtraction on the $C_{n}^{2}$ profile
      reconstruction. Left: $30 \times 30$ geometry. Right: $15 \times 15$
      geometry.}
    \label{fig:impact_soustraction_biais}     
  \end{center}
\end{figure}

Then, we carry out the bias subtraction, as explained in Subsection
\ref{sub:bias_subtraction}, and using $\sigma_{\Delta \phi_{noise}}^{2}$ and
$\sigma_{\delta i_{noise}}^{2}$ values presented in Table
\ref{tab:SNR_slo} and \ref{tab:SNR_sci}. They are subtracted from the central
value of the auto-correlation maps. Bias-subtracted correlations are then used
to restore the $C_{n}^{2}$ profile, and results are presented in
Fig. \ref{fig:impact_soustraction_biais}. 
This correction nearly allows to restore the unbiased profile, especially
with the $30 \times 30$ geometry. Slight differences can be noticed with the
$15\times 15$ geometry, probably due to the poor SNR, but the efficiency of
the method is demonstrated.

\subsection{Comparison with other methods,  advantages of the two SH geometries}\label{sub:comparisons}

\begin{figure}[!h]
  \begin{center}
    \resizebox{0.9\columnwidth}{!}{
      \includegraphics{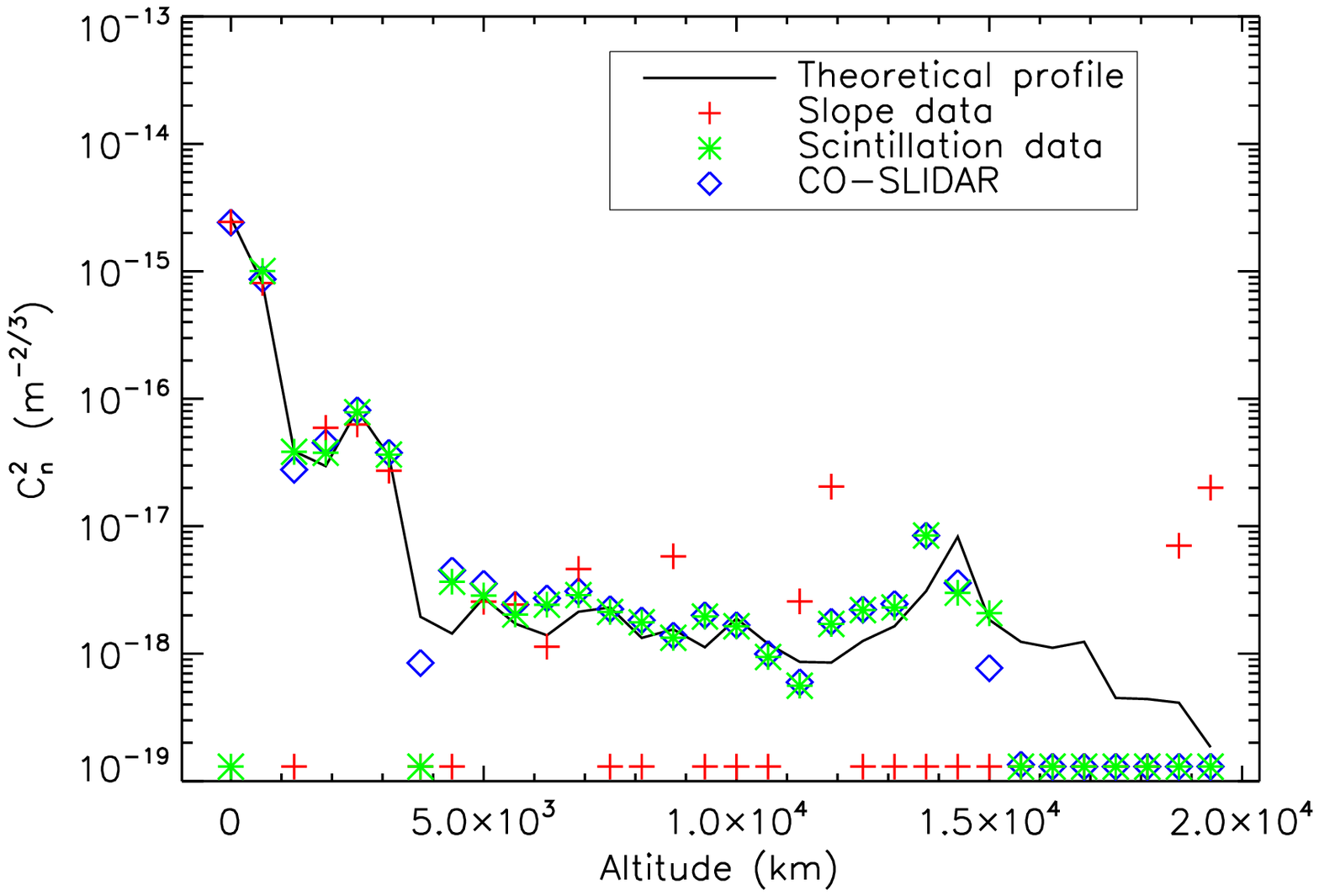}
      \includegraphics{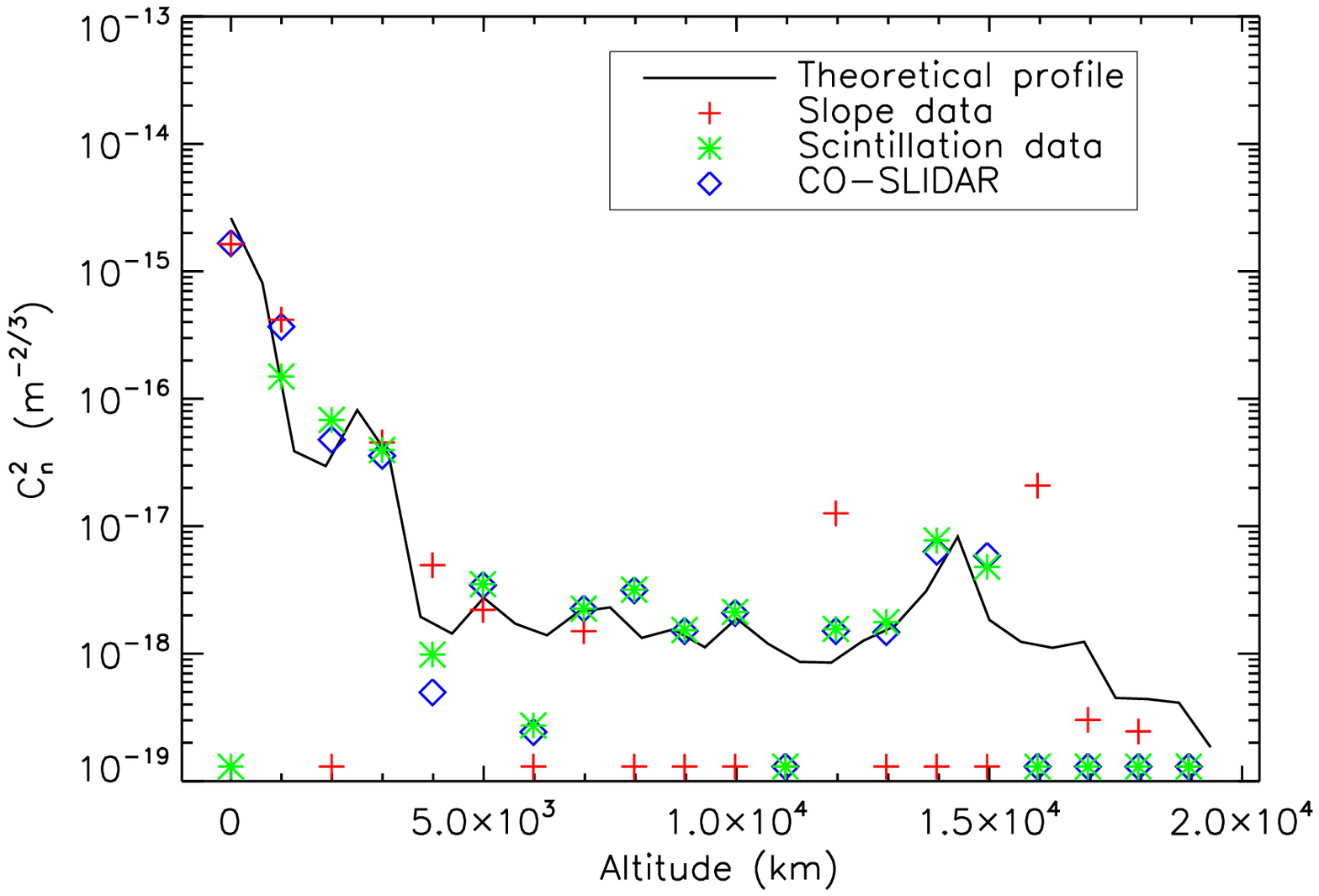}}
    \caption{Comparison of CO-SLIDAR with other methods. The bias has been
      subtracted in the three cases. Left: $30 \times 30$
      geometry. Right: $15 \times 15$ geometry}
    \label{fig:comp_methodes}     
  \end{center}
\end{figure}

We finally compare $C_{n}^{2}$ profiles restored with CO-SLIDAR with $C_{n}^{2}$
profiles restored with slope data only or scintillation data only. Results are
shown in Fig. \ref{fig:comp_methodes}. 
Comments are common to both
geometries. In all cases, noisy data were used,
and the bias has been calculated and subtracted. Slope data allow a good reconstruction
of ground and low altitude layers, but they provide worse sensitivity at high
altitude. Scintillation data perform a good reconstruction at low and high
altitude, but do not permit to estimate the ground layer, as there is no
scintillation on the pupil. This goes in the way of using both type of data
to retrieve the $C_{n}^{2}$ profile. Actually, CO-SLIDAR combined the
advantages of the two kind of data in a single instrument, and allows to
estimate an accurate $C_{n}^{2}$ profile from the ground to $15~km$ of
altitude, with these SH geometries. Layers after $15~km$ of altitude are not
well reconstructed, neither with CO-SLIDAR nor slope nor scintillation data
only. Indeed, after $15~km$, with these geometries, cross-correlations are
blind to turbulence, according to Eq. \ref{eq:alt_max}. Information can then be
provided by scintillation auto-correlations, but here the scintillation
signal seems to be too weak to perform a good estimation of $C_{n}^{2}$
values. Better sensitivity could be achieved using smaller
subapertures\cite{2012aoeltjv}, but this would lead to deal with very low fluxes at
subaperture level in an astronomical context. Anyway, these $C_{n}^2$ values are very small, $\le 1
\times 10^{-18}~m^{-\frac{2}{3}}$, and represent a negligible part of the
whole turbulence. Nevertheless, these layers over $15~km$ of altitude could be
estimated using a binary star with
a smaller separation, to increase the altitude range sensitivity, but this would
decrease the altitude resolution.

\begin{figure}[!h]
  \begin{center}
    \resizebox{0.7\columnwidth}{!}{
      \includegraphics{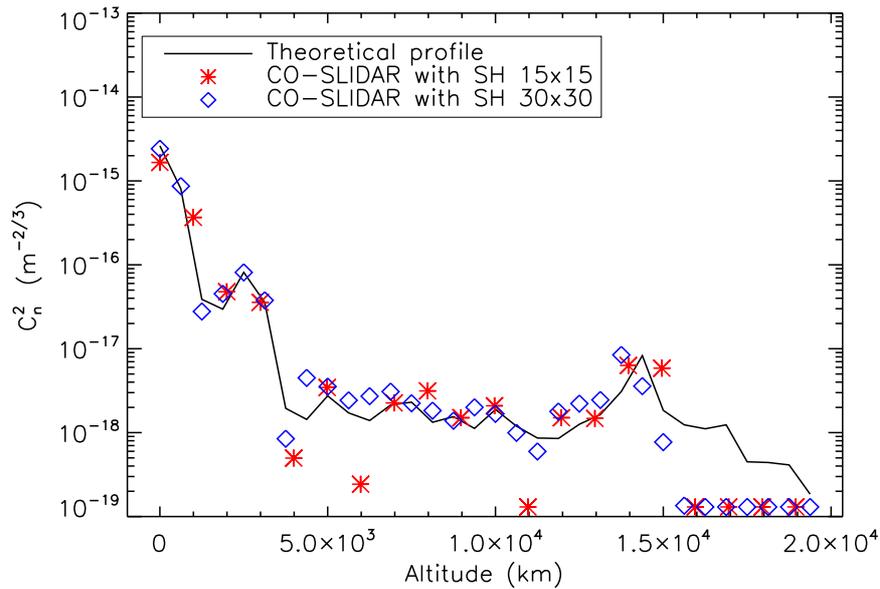}}
    \caption{Comparison of the two SH geometries.}
    \label{fig:comp_geom}     
  \end{center}
\end{figure}

Finally, the performances of the two SH geometries need to be compared. Reconstructed
profiles are presented in Fig. \ref{fig:comp_geom}. 
We immediately notice the $30\times 30$ geometry has a better altitude resolution, about
$500~m$, while the $15\times 15$ geometry only permits a resolution of about
$1~km$, as forecast by Eq. \ref{eq:resolution}. Both SH geometries allows a $C_{n}^{2}$ profile
measurement from the ground to $15~km$ but do not permit to go over with
reliable results, as explained before. The $15\times 15$ geometry, by
collecting more flux, allows a better sky-coverage. Nevertheless, one has to
be careful because if turbulence is too strong, images will be formed of
speckles, leading to a difficult extraction of slopes and scintillation
indexes. Both geometries have their advantages and they could be used together
in order to cover a full observation night.

\section{Conclusions and perspectives}\label{sect:conclusion}

In this paper, we detailed the performances of CO-SLIDAR in an end-to-end
simulation, in a realistic astronomical case. Detection noises have been taken
into account, and a way to subtract the bias from the measurements has been
proposed and its efficiency demonstrated. CO-SLIDAR has been compared to other methods, highlighting the good
complementarity of correlations of slopes and scintillation indexes, to
provide turbulence sensitivity at low and high altitude. Two SH geometries
have been proposed, each one with its proper advantages.

In order to perform complete on-sky validation, we acquired CO-SLIDAR data
through an observation campaign on the MeO $1.5$-meter telescope, with the two
SH geometries, at the Côte d'Azur Observatory, in South of France. Data processing is in
progress, in order to estimate high-resolution $C_{n}^{2}$ profiles.
SCO-SLIDAR, the extension of the method to a single source, should also be
tested on single star data. Computation of error bars on the reconstructed
profile has to be implemented, together with a refined study of the convergence
noise. As sensitivity to the outer scale $L_{0}$ has
been proved in a previous work\cite{2012aoeltjv}, combined estimation of the
two parameters is conceivable.   

\section{Acknowledgements}

This work has been performed in the framework of a Ph.D Thesis supported by
Onera, the French Aerospace Lab, and the French Direction Générale de
l'Armement (DGA).

\end{document}